# Visible spectra of $W^{8+}$ in an electron-beam ion trap

Q. Lu (陆祺峰),[1] C. L. Yan (严成龙),[1] J. Meng (孟举),[2] G. Q. Xu (许帼芹),[1] Y. Yang (杨洋),[1] C.Y. Chen (陈重阳),[1] J. Xiao (肖君),[1,*] J.G. Li (李冀光),[2,†] J.G. Wang (王建国),[2] and Y. Zou (邹亚明)[1]

[1] Shanghai EBIT Laboratory, Key Laboratory of Nuclear Physics and Ion-Beam Application (MOE), Institute of Modern Physics, Fudan University, Shanghai 200433, China;

[2] Institute of Applied Physics and Computational Mathematics, Beijing 100088, China

Corresponding author: *xiao_jun@fudan.edu.cn, †Li_Jiguang@iapcm.ac.cn

To provide spectroscopic data for lowly charged tungsten ions relevant to fusion research, this work focuses on the $W^{8+}$ ion. Six visible spectra lines from $W^{8+}$ in the range of 420–660 nm are observed with a compact electron-beam ion trap in Shanghai. Furthermore, transition energies are calculated for the 30 lowest levels of the $4f^{14}5s^25p^4$, $4f^{13}5s^25p^5$ and $4f^{12}5s^25p^6$ configurations of $W^{8+}$ by using the flexible atomic code (FAC) and GRASP package, respectively. Reasonably good agreement is found between our two independent atomic-structure calculations. The resulting atomic parameters are adopted to simulate the spectra based on the collisional-radiative model implemented in the FAC code. This assists us with identification of six strong magnetic dipole transitions in the $4f^{13}5s^25p^5$ and $4f^{12}5s^25p^6$ configurations from our experiments.

## I. INTRODUCTION

As one of potential candidates for plasma-facing material in tokamaks [1-3], tungsten (W) is considered to be the main impurity ions in the ITER plasma, which could give rise to undesirable radiative power losses. To assess W flux rates and diagnose plasma, numerous spectroscopic data for W ions are demanded [4]. Moreover, W can be ionized up to $W^{64+}$ in tokamak plasma due to the wide range of electron temperatures, from 0.1 keV at the edge to 20 keV in the core [5]. Therefore, many experimental and theoretical studies have been carried out by different groups to provide atomic parameters (see, e.g., Refs. [8-40], and references therein). However, according to the W data compiled by Kramida *et al*. [8-10], there is still short of spectroscopic data for lowly charged W ions, e.g., $W^{7+}$–$W^{26+}$, which calls for more spectra measurements and accurate line identifications.

Electron-beam ion traps (EBITs) are one of the most versatile light sources for studying W spectroscopy (see, e.g., Refs. [20,36-40]), since W ions in almost any charge state can be produced through successive ionization by a monoenergetic and energy-adjustable electron beam. However, the determination of charge states of lowly charged W ions is an intractable problem in EBIT measurements (see, e.g., Refs. [26,27,41]). This difficulty is caused by a couple of reasons. Firstly, possible metastable levels in certain charged ions may result in the so-called indirect ionization, that is, early production of W ions despite the lower electron-beam energy than the corresponding ionization potential [36,38,42,43]. Secondly, the interval between ionization potential of adjacent lowly charged W ions is small but the uncertainties of their ionization potentials are large due to intricate electron correlations. Last but not least, multi-charged W ions have a few opened shells, especially the *f* orbital, which brings about the complicated spectra [44,45].

It is also a challenging work for atomic-structure calculations to predict accurate atomic parameters. The many-body perturbation theory (MBPT) cannot be applied to deal with strong electron correlation effects in lowly charged W ions, so the configuration interaction (CI) method has to be adopted [44]. For the reliable CI calculations, one should include a large number of configurations to capture major electron correlations, but this requires very large computational resource [45].

Over the last few years some progress has been made in spectroscopic data for lowly charged W ions (gap for $W^{7+}$–$W^{26+}$), including $W^{7+}$ [4,33,34,36,46,47,48], $W^{11–15+}$ [24-27], and $W^{25–26+}$ [37,39]. In the present work we focus on the visible spectra lines from $W^{8+}$, a more complicated case than $W^{7+}$. Prior experimental studies were conducted by Ryabtsev *et al*. [46] and Mita *et al*. [47]. Ryabtsev *et al*. discovered 483 lines of $W^{8+}$ in the extreme ultraviolet (EUV) region via vacuum spark sources, but the identification of measured lines were not accomplished probably because of the large uncertainty of theoretical calculations for $W^{8+}$ at that time. Using an EBIT device, Mita *et al*. found five visible lines and an EUV array of lines in $W^{8+}$. The EUV lines from $W^{8+}$ were then identified [48] with assistance of a spectra simulation based on the collision radiative model, though the deviation of wavelengths is more than 5% from experiments on average. However, the identification of those visible lines for $W^{8+}$ remains unsolved. The atomic structure for $W^{8+}$ is very complicated since the orbital energies of the $4f$ and $5p$ orbitals are almost the same. Thus, energy levels $4f^{13}5s^25p^5$ $^3F_4$ and $4f^{14}5s^25p^4$ $^3P_2$ compete for the ground state of $W^{8+}$, as pointed out by Kramida and Shirai [8]. They predicted the ground and first excited state of $W^{8+}$ based on atomic-structure calculations by Cowan's codes [8]. For searching for proof of time variation of fine structure constant $\alpha$, Berengut *et al*. calculated excitation energies of the lowest 16 levels for $W^{8+}$ employing the MBPT+CI method [49]. More recently, the transition energies and decay rates of states in $W^{8+}$ below the first ionization threshold were studied theoretically by Kozioł and Rzadkiewicz employing the multiconfiguration Dirac–Hartree–Fock (MCDHF) and the relativistic CI methods [50]. However, results from these three independent calculations are different from each other to a large extent, for example, a 8000 cm$^{-1}$ difference in the excitation energy of the first excited level. To sum up, more systematic studies are required to provide reliable spectroscopic data for $W^{8+}$.

In this work, we use the Shanghai high temperature superconducting electron-beam ion trap (SH-HtscEBIT) [51] to re-measure the visible spectra of the $W^{8+}$ ion. At the same time, we calculate the transition energies for the lowest 30 levels of $W^{8+}$ employing the flexible atomic code (FAC) [52] and the GRASP package [53,54]. Furthermore, the measured lines are identified with aid of spectra simulation based on the collisional-radiative model (CRM) implemented in FAC.

## II. EXPERIMENTS

The present experimental devices and procedure are essentially the same as those in our previous studies [24-26,36-38], so here we only give a brief description. The SH-HtscEBIT is specially designed for doing researches related to lowly charged ions. It mainly consists of an electron gun, three drift tubes (DT1, DT2, and DT3), an electron collector, a high-temperature superconducting coil, a liquid nitrogen tank, and a gas injection system. The electron beam emitted from the electron gun is accelerated toward the drift tubes and its radius is compressed to ~150 μm by the magnetic field produced by the superconducting coil operating at liquid-nitrogen temperature. Once the electron beam reaches the drift tube region, it collides with atoms from the injection system to produce targeted ions, whose charge state mainly depend on the energy difference between DT2 and cathode of electron gun. Finally, the electrons are collected by the electron collector.

In this work, W(CO)$_6$ is chosen to produce the W$^{8+}$ ions via the gas injection system, since it is a volatile compound which has a high enough vapor pressure at room temperature. Once the targeted ions are formed, they are confined radially by both the space charge potential of electrons and the magnetic field, while axially confined by the potential well (~100 V). The fluorescence emitted from the trapped ions is observed by a Czerny-Turner spectrometer (Andor Sr-303i) covering the range of 200–800 nm and then fitted with a 1200 L/mm grating blazed at 500 nm. To obtain a larger collection solid angle, a convex lens of $f$ =150 mm is used to focus light from the trap on the entrance slit of spectrometer, which is set to 30 μm. Finally, the photons are detected by an EMCCD camera (Andor DU971P-UVB) operated at -65 °C.

### III. THEORETICAL CALCULATIONS

As discussed in previous studies [26,37,39,40,55-57], the strong visible lines observed in Tokamak or EBIT plasma mostly come from the low-lying states of the targeted ions, since these states are largely populated. Thus, in this work we focus on the 30 energy levels belonging to the three lowest configurations $4f^{14}5s^25p^4$, $4f^{13}5s^25p^5$ and $4f^{12}5s^25p^6$.

**A. FAC calculations**

The popular FAC code (version 1.1.5) is used to provide the atomic data [52]. Two theoretical methods including the relativistic CI (RCI) and MBPT (RMBPT) are implemented in FAC. Note that only the second-order perturbation correction is made for the latter, which is not applicable to atomic systems that have strong electron correlation effects such as the case under investigation. Here RCI method is selected to calculate the complex energy structure for W$^{8+}$.

The RCI method is widely used due to its high-efficiency and credible predictions for experimental findings [52,58-61]. As a starting point, it constructs a fictitious mean configuration with fractional occupation numbers that takes into account the electron screening of the involved configurations. The orbitals are optimized self-consistently in the Dirac-Hartree-Fock-Slater (DHFS) approximation to minimize the average energy of the fictitious mean configuration. The atomic state function (ASF) is composed of configuration state functions (CSFs), which are antisymmetrized linear combinations of products of single-electron Dirac orbitals. By solving the eigenvalue problems, mixing coefficients and level energies can be obtained.

In our work, to better calculate the lowest 30 levels of W$^{8+}$, we include $4f^{12}5s^25p^6$, $4f^{14}5s^25p^4$ and $4f^{13}5s^25p^5$ to form the basis for the construction of mean configuration. Next, we do large-scale calculations to capture the main electron correlations. Trial calculations are conducted in order to select important configurations that have large influence on the excitation energies of the lowest 30 levels. To balance the computational resources and the calculation accuracy, only those configurations that have the influence over 1% on any of the 30 energy levels are included in our final calculation. In detail, single and double excitation from the 4$f$, 5$s$ and 5$p$ electron of $4f^{14}5s^25p^4$, $4f^{13}5s^25p^5$ and $4f^{12}5s^25p^6$ to $n$=5 and $n$=6 are considered in our calculation. Additionally, the single excitation of 4$f$ electron from $4f^{13}5s^25p^5$ and $4f^{12}5s^25p^6$ to 7$f$, 8$f$, 9$f$ and 10$f$ are taken into account, which contributes to the excitation energies of the lowest 30 energy levels for W$^{8+}$ by approximately 10%. Moreover, it is found that the correction from the 4$d$ electron correlation reaches 20% to the excitation energies. Thus, some single and double excitation from the 4$d$ electron are also included, involving $4d^94f^{14}5s^25p^45d$, $4d^94f^{14}5s^25p^45f$, $4d^94f^{13}5s^25p^55d$, $4d^94f^{13}5s^25p^55f$, $4d^94f^{13}5s^25p^56s$, $4d^94f^{13}5s^25p^56p$,

$4d^94f^{13}5s^25p^56d$, $4d^94f^{13}5s^25p^56f$, $4d^94f^{12}5s^25p^65d$, $4d^94f^{12}5s^25p^65f$, $4d^94f^{12}5s^25p^66s$, $4d^94f^{12}5s^25p^66p$, $4d^94f^{12}5s^25p^66d$, $4d^94f^{12}5s^25p^66f$, $4d^84f^{14}5s^25p^6$, $4d^84f^{14}5s^25p^45d^2$, $4d^84f^{14}5s^25p^45d5f$, $4d^84f^{14}5s^25p^45f^2$, $4d^84f^{13}5s^25p^55d^2$, $4d^84f^{13}5s^25p^55d5f$, and $4d^84f^{14}5s^25p^45f^2$. Furthermore, several small corrections arising from Breit interaction, vacuum polarization and electron self-energy are considered. Finally, 537988 energy levels are obtained based on the above consideration.

### B. GRASP calculations

To check the reliability of our results obtained with the FAC code, we also perform the MCDHF calculation by using the GRASP package [53,54]. The MCDHF method is more efficient to capture electron correlations than RCI. Moreover, no approximation is made in the Dirac-Hartree-Fock potential. We start from the self-consistent field (SCF) calculation, in which all orbitals occupied in configurations $4f^{12}5s^25p^6$, $4f^{13}5s^25p^5$, and $4f^{14}5s^25p^4$ are optimized as spectroscopic orbitals. Here, we omit the common core $1s^22s^22p^63s^23p^63d^{10}4s^24p^64d^{10}$ orbitals in the notation of the configurations for convenience. Electron correlation effects can be captured by configuration state functions generated through single and double excitations from occupied orbitals in the multi-configuration set to virtual ones. The correlation between electrons in the core with $n = 4$ and 5 and those in the valence subshells is included together with the valence-valence correlation between electrons in the 5$p$ and 4$f$ subshells and the core-core (CC) electron correlation in the 5$s$ subshell. It means that we produced CSFs by replacing these orbitals with virtual ones. A restriction is made so that only one electron can be promoted from orbitals in the 5$p$ and 4$f$ core subshells at a time. Virtual orbitals are augmented layer by layer up to $n = 7$ and $l = 5$. To avoid convergence issue, only the virtual orbitals in the last added layer are variable in SCF procedures. Additionally, we remove all off-diagonal elements of the Hamiltonian matrix except for those interacting with the multi-configuration set in the SCF calculations. This approximation is equivalent with the second-order perturbation theory. However, it should be emphasized that those neglected off-diagonal matrix elements are included in the subsequent RCI computation. Also, the Breit interaction and QED corrections are considered in this step. It is found that the CC correlation between the 5$s$ electrons play a nonnegligible role in the excitation energies concerned.

### C. Results from FAC and GRASP calculations

The excitation energies for the lowest 30 states of W[8+] from our FAC and GRASP calculations are listed in Table I. We also present the theoretical results by Kozioł and Rzadkiewicz [50] and by Berengut *et al*. [49]. The differences in excitation energies between our FAC and GRASP calculations and previous theories [49,50] are illustratively shown in Fig. 1 as well.

As can be seen from Table I, our results confirm that the ground state is $4f^{14}5s^25p^4$ $^3P_2$ for W[8+], and the excitation energy is around 12000 cm[-1] for the first excited state $4f^{13}5s^25p^5$ $^3F_4$. Overall, the excitation energies agree between our two independent calculations, and the deviation is ~1.5% on average. The largest discrepancies between our FAC and GRASP calculations are less than 5000 cm[-1] for levels from $4f^{12}5s^25p^6$. In contrast, the excitation energies reported by Kozioł and Rzadkiewicz overall agree with our results for levels from $4f^{14}5s^25p^4$ and $4f^{13}5s^25p^5$, but differ from our calculations a lot for levels from $4f^{12}5s^25p^6$, while those from Berengut *et al*. are systematically smaller than our results by 6000 cm[-1].

TABLE I. Lowest 30 energy levels of $W^{8+}$ given by our FAC and GRASP calculations. Part of results presented by Kozioł and Rzadkiewicz [50] and by Berengut *et al*. [49] are also added. $E$ (in cm$^{-1}$) represents level energies relative to the ground state. Note, results from other theoretical methods that do not correspond to the order given by our FAC calculation are highlighted in boldface.

| | | | $E$ (cm$^{-1}$) | | | |
|---|---|---|---|---|---|---|
| No. | Configuration | $J$ | FAC | GRASP | Kozioł | Berengut |
| 1 | $4f^{14}5s^25p^4$ $^3P$ | 2 | 0 | 0 | 0 | 0 |
| 2 | $4f^{13}5s^25p^5$ $^3F$ | 4 | 11856 | 12128 | 9440 | 6075 |
| 3 | $4f^{13}5s^25p^5$ $^1F$ | 3 | 12412 | 12219 | 10445 | 6357 |
| 4 | $4f^{13}5s^25p^5$ $^3G$ | 5 | 16762 | 16135 | 14767 | 11122 |
| 5 | $4f^{14}5s^25p^4$ $^1S$ | 0 | 25411 | 25113 | **26393** | **29810** |
| 6 | $4f^{13}5s^25p^5$ $^3F$ | 3 | 27219 | 27369 | **24791** | **21905** |
| 7 | $4f^{13}5s^25p^5$ $^1D$ | 2 | 28739 | 27553 | 26847 | **23276** |
| 8 | $4f^{13}5s^25p^5$ $^3F$ | 2 | 33621 | 33081 | 31473 | **28112** |
| 9 | $4f^{13}5s^25p^5$ $^3G$ | 4 | 40051 | 39361 | 38190 | 34884 |
| 10 | $4f^{13}5s^25p^5$ $^3D$ | 1 | 42194 | 40637 | 40417 | 36497 |
| 11 | $4f^{12}5s^25p^6$ $^3H$ | 6 | 64401 | 69054 | 56443 | 56416 |
| 12 | $4f^{12}5s^25p^6$ $^3F$ | 4 | 72626 | 77429 | 65554 | 65008 |
| 13 | $4f^{12}5s^25p^6$ $^3H$ | 5 | 80687 | 85311 | 72409 | 73188 |
| 14 | $4f^{14}5s^25p^4$ $^3P$ | 1 | 86109 | 86791 | **85630** | |
| 15 | $4f^{12}5s^25p^6$ $^3H$ | 4 | 87655 | 92375 | **79885** | **80551** |
| 16 | $4f^{12}5s^25p^6$ $^3F$ | 2 | 89430 | 93125 | 83684 | **82424** |
| 17 | $4f^{12}5s^25p^6$ $^3F$ | 3 | 90495 | 94886 | **83361** | **83315** |
| 18 | $4f^{14}5s^25p^4$ $^1D$ | 2 | 100042 | 100942 | **101530** | |
| 19 | $4f^{12}5s^25p^6$ $^1G$ | 4 | 103109 | 108056 | **95161** | |
| 20 | $4f^{13}5s^25p^5$ $^3D$ | 3 | 111500 | 111896 | **110930** | |
| 21 | $4f^{12}5s^25p^6$ $^3P$ | 2 | 112224 | **115685** | 106798 | |
| 22 | $4f^{13}5s^25p^5$ $^1G$ | 4 | 114282 | **115049** | 113533 | |
| 23 | $4f^{12}5s^25p^6$ $^1I$ | 6 | 122996 | 124065 | 116028 | |
| 24 | $4f^{12}5s^25p^6$ $^3P$ | 0 | 124162 | **127363** | 120388 | |
| 25 | $4f^{13}5s^25p^5$ $^3G$ | 3 | 124769 | **125712** | **124043** | |
| 26 | $4f^{12}5s^25p^6$ $^3P$ | 1 | 127441 | 130060 | **122864** | |
| 27 | $4f^{12}5s^25p^6$ $^3P$ | 2 | 130901 | 133694 | 126048 | |
| 28 | $4f^{13}5s^25p^5$ $^1D$ | 2 | 138479 | 138589 | 132931 | |
| 29 | $4f^{12}5s^25p^6$ $^1S$ | 0 | 179372 | 183515 | 187738 | |
| 30 | $4f^{14}5s^25p^4$ $^3P$ | 0 | 206478 | 210873 | 201003 | |

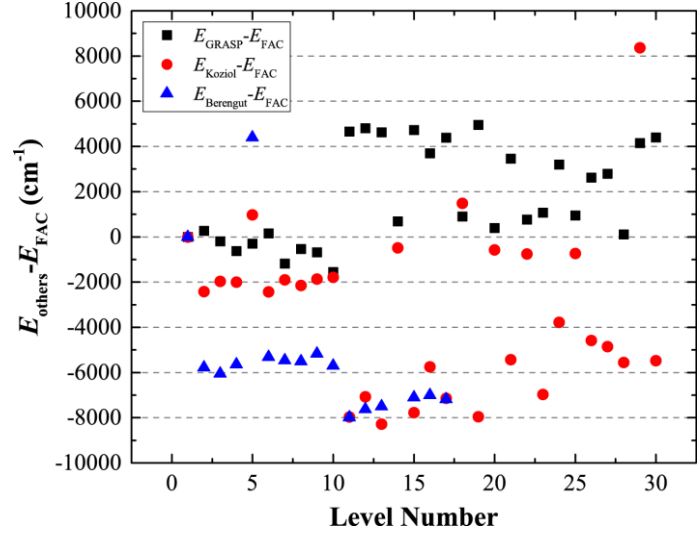

FIG. 1. Comparisons of the excitation energies of our GRASP calculations (black squares), Kozioł and Rzadkiewicz's calculations (red circles), and Berengut *et al*. (blue triangles) with our FAC results.

## IV. SPECTRAL MEASUREMENTS AND LINE IDENTIFICATIONS
### A. Charge state determination of spectral lines

We measure spectra from the lowly charged W ions at nominal electron-beam energies of 100, 120, 140 and 160 eV in visible region ranging from 300–700 nm. Here, we present the measured spectra between 420 and 660 nm in Fig. 2 where the visible spectral lines from the $W^{8+}$ ion are found. To obtain the actual electron-beam energy, we adopt similar treatments as in our previous studies (see Ref. [36,62,63] for details), since the plasma environments are almost the same.

Wavelength calibration is performed in the W spectra measurement intervals using emission lines from Pen-Ray lamps, and the wavelengths of reference lines are taken from the NIST [64]. The dispersion function from the CCD pixel number to the wavelength is obtained by using a quadratic polynomial fitting. In this work, the fitting uncertainties, mainly due to the low signal-to-noise ratio, are 0.003–0.008 nm. The dispersion function brings about the 0.001–0.041 nm uncertainties for measured lines, depending on wavebands and the lines' positions on CCD. Systematic uncertainties, mainly caused by the difference between the positions of the lamps and ions, are estimated to be 0.02nm. Considering the above three uncertainties, the total uncertainties of our wavelength calibration are 0.02–0.05 nm.

As can be seen from Fig. 2, numerous lines are found in the measured range. Some of these lines (indicated by the black brackets) are from lowly-charged background ions such as nitrogen and oxygen, which also exist in the spectra without injection. After excluding the background lines, we classify these lines into several groups according to their intensity variation as a function of electron-beam energy. One line at 574.45(3) nm comes from the $W^{7+}$ ions [36,47]. Six lines marked as A1–A6 are assigned to the $W^{8+}$ ion, which confirm the previous observation reported in Ref. [47]. It is worth noting that the A6 line is observed for the first time. In addition, three lines at 438.63(4), 481.52(3), and 608.37(5) nm are connected to $W^{9+}$. We should emphasize that these lines from the $W^{7+}$, $W^{8+}$, and $W^{9+}$ ions appear at electron-beam energies far below their corresponding ionization energies [64]. This indicates the common existence of the indirect ionization in lowly charged W ions.

The wavelengths as well as the uncertainties of A1–A6 are listed in Table II. For comparison, we also give in the third column of this table the wavelengths deduced from Fig. 1 in Ref. [47]. Our experimental results for A1–A5 are consistent with Mita's within the error bars. We also convert the air wavelengths to the corresponding vacuum wavelengths for comparing with theoretical calculations that are displayed in the fourth column of Table II. The fifth column is the corrected intensity for each line at electron-beam energy 104.5 eV according to the efficiency curves of our instruments, since the efficiencies of the grating and the CCD are not constant at different wavelengths. The last column is the normalized intensities to A5.

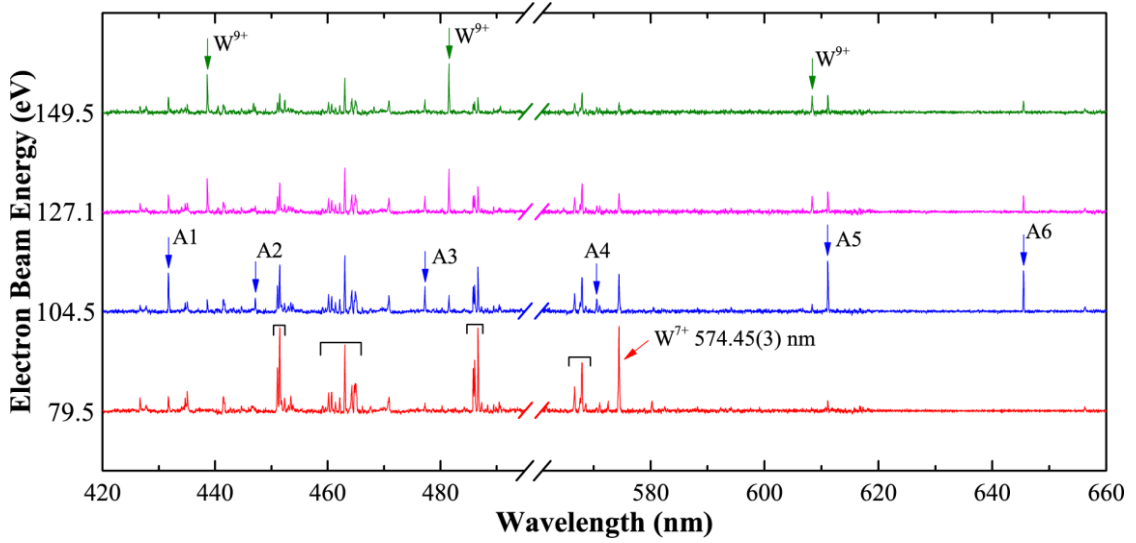

FIG. 2. Spectra from lowly charged W ions obtained by SH-HtscEBIT at corrected electron-beam energy of 79.5, 104.5, 127.1 and 149.5 eV in the range of 420–660 nm. Accumulation time of each spectrum is 3 hours. Six arrowed lines A1–A6 are assigned to come from the $W^{8+}$ ion, while line at 574.45(3) nm is the M1 transition between the fine structure levels in the $4f^{13}5s^25p^6$ $^2F$ ground term of $W^{7+}$.

TABLE II. Information of six spectral lines from our measurements. $\lambda_{\text{exp}}^{\text{air}}$ and $\lambda_{\text{exp}}^{\text{vac}}$ represent the wavelengths in air and vacuum respectively, and the units are given in nm. $\lambda_{\text{Mita}}^{\text{air}}$ is the deduced wavelength (air) of each $W^{8+}$ line in Ref. [47]. The experimental uncertainties for wavelength are given in the parentheses. $I_{\text{corr}}$ is the intensity of each observed line at electron-beam energy 104.5 eV after efficiency correction. $I_{\text{nom}}$ is the normalized intensity to A5.

| Key | $\lambda_{\text{exp}}^{\text{air}}$ | $\lambda_{\text{Mita}}^{\text{air}}$ | $\lambda_{\text{exp}}^{\text{vac}}$ | $I_{\text{corr}}$ | $I_{\text{nom}}$ |
|---|---|---|---|---|---|
| A1 | 431.73(5) | 431.9(3) | 431.84 | 3591 | 0.71 |
| A2 | 447.13(2) | 447.4(3) | 447.25 | 1210 | 0.24 |
| A3 | 477.25(3) | 477.3(3) | 477.38 | 1962 | 0.39 |
| A4 | 570.56(4) | 570.6(3) | 570.72 | 1206 | 0.24 |
| A5 | 611.13(2) | 610.7(5) | 611.30 | 5062 | 1.00 |
| A6 | 645.48(2) |  | 645.66 | 4656 | 0.92 |

### B. Line identification by CRM simulation

For unambiguous line identifications, a mere match between measured and calculated wavelengths is not sufficiently rigorous due to dense spectra lines and computational uncertainties of atomic parameters. Thus, the comparisons of line intensities are also significant. In an optically thin plasma like EBIT, the intensity of a spectral line from the upper level $i$ to lower level $j$ can be defined as:

$$I_{i,j}(\lambda) \propto N_i A_{i,j} \phi(\lambda),$$

where $\lambda$ is the wavelength and $A_{i,j}$ is the transition probability. The function $\phi(\lambda)$ is the normalized line profile, which is taken as a Gaussian profile to include contributions from Doppler, natural, collisional and instrumental broadenings. $N_i$ is the population of upper level, which is determined by the interactions with plasma particles.

CRM is widely used to aid line identification in the observed spectrum (see, e.g., Refs. [2,20,22], and references therein). In the framework of CRM, all physical processes are supposed to be taken into account to build and solve a system of rate equations for level populations. As a compromise between computational accuracy and efficiency, however, only electron-impact excitation, electron-impact deexcitation, and radiative decay are included in our CRM simulation. Other dynamical physical processes such as dielectronic recombination, radiative recombination, three-body recombination, charge exchange, etc. are omitted in this work, since their influences are estimated to be small in our experiment [36-38,41]. We can use the following equation to describe the differential rate of the population of each energy level:

$$\frac{dN_i}{dt} = \sum_{j>i}\left(A^r_{j \to i} N_j\right) + \sum_{j<i}\left(C^e_{j \to i} N_j n_e\right) + \sum_{j>i}\left(C^d_{j \to i} N_j n_e\right)$$
$$- \sum_{j<i}\left(A^r_{i \to j} N_i\right) - \sum_{j>i}\left(C^e_{i \to j} N_i n_e\right) - \sum_{j<i}\left(C^d_{i \to j} N_i n_e\right),$$

where $N$ is the population of each level, the subscripts $(i, j)$ represent the initial and final energy levels, and $n_e$ denotes the electron density of plasma. $C^e$ and $C^d$ stand for electron-impact excitation and deexcitation rate coefficient, respectively, which can be obtained by convoluting the cross section of the electron-impact excitation (deexcitation) with the free electron energy distribution function (Gaussian function in our case). Cross sections of electron-impact excitations are calculated by the distorted wave approximation in FAC, while the cross section of the electron-impact deexcitation can be then calculated according to the principle of detailed balance. Considering quasi-steady-state approximation $\frac{dN_i}{dt} = 0$ and normalized condition $\sum_i N_i = 1$, we can solve the equation above and obtain the population of each energy level. Finally, the intensity of each transition is obtained and strong lines are selected for comparisons with experiments.

Note that the CRM simulation requires very large computer memory resources to solve a large number of rate equations, we conduct a small-scale RCI calculation using FAC package to provide necessary atomic data for CRM simulation, which is accomplished to the best of our computer resources. A total of 28903 energy levels are obtained by considering important configurations including $4f^{14}5s^25p^4$, $4f^{13}5s^25p^5$, $4f^{12}5s^25p^6$, $4f^{13}5s^25p^4nl$ ($n$=5, $l$=$d$, $f$, $g$), $4f^{13}5s^25p^4nl$ ($n$=6, $l$=$s$, $p$, $d$, $f$), $4f^{14}5s5p^4nl$ ($n$=5, $l$=$p$, $d$, $f$, $g$), $4f^{14}5s5p^4nl$ ($n$=6, $l$=$s$, $p$, $d$, $f$), $4f^{14}5p^6$, $4f^{14}5s^25p^3nl$ ($n$=5, $l$= $d$, $f$, $g$), $4f^{14}5s^25p^3nl$ ($n$=6, $l$=$s$, $p$, $d$, $f$), $4f^{12}5s^25p^5nf$ ($n$=5–10), $4f^{13}5s5p^55l$ ($l$=$p$, $d$), $4f^{13}5s5p^56l$ ($l$=$s$, $d$), $4f^{11}5s^25p^6nf$ ($n$=5–10), $4f^{12}5s5p^65d$, $4f^{14}5p^45ll'$ ($l$, $l'$=$d$, $f$), $4f^{14}5p^55l$ ($l$=$d$, $f$), $4f^{14}5p^56l$ ($l$=$s$, $p$, $d$, $f$), $4f^{14}5p^45d6l$ ($l$=$s$,$p$,$d$,$f$), $4f^{14}5p^45f6l$ ($l$=$s$,$p$,$d$,$f$), $4f^{13}5p^65l$ ($l$=$d$,$f$), $4f^{13}5p^55ll'$ ($l$, $l'$=$d$, $f$), $4f^{13}5p^66l$ ($l$=$s$,$p$,$d$,$f$), $4f^{13}5p^55d6l$ ($l$=$s$,$p$,$d$,$f$), $4f^{13}5p^55f6l$ ($l$=$s$,$p$,$d$,$f$), $4f^{12}5p^65ll'$ ($l$, $l'$=$d$,$f$), $4f^{12}5p^65d6f$ and $4f^{12}5p^65f6f$. For transition data, E1, M1, E2, and M2 transitions within the $n$=5 complex are calculated, while higher-order transitions such as E3 and M3 are omitted due to their small contributions. In addition, E1 transitions from $n$=6 complex to the lower complex are also considered. For electron-impact (de)excitation rate, only configurations within the $n$=5 complex are included. To match the experimental settings, our CRM is conducted at electron density $5 \times 10^{10}$ cm$^{-3}$ and electron-beam energy 104.5 eV. Six strongest lines in the range of 420–660 nm according to our CRM calculation are found. By comparing both the wavelengths ($\lambda_{\text{FAC}}^{\text{a}}$) and intensities ($I_{\text{CRM}}$) with the experimental results, we assign them to the six observed lines A1–A6, and the results, including calculated wavelengths and transition rates, are shown in Table III.

As can be seen from Table III, all the theoretically predicted strong lines are observed in our EBIT measurements, indicating our CRM simulations are reasonable to some extent. Lines A1, A3 and A6 are identified to magnetic-dipole (M1) transitions within the $4f^{13}5s^25p^5$ configuration, while A2, A4 and A5 come from M1 transitions within the $4f^{12}5s^25p^6$ configuration. Our calculated wavelengths of six lines deviate from our experimental results by 1.38% on average. This difference is acceptable because only limited configurations are included in our small-scale CRM calculation. For comparisons of intensity, the results from our experiments and CRM simulation are shown in Fig. 3. It is found that the CRM and experimental data points from A1 to A6 exhibit similar trend. In more detail, both our experiments and calculation show that A2–A4 are the three weakest lines among the six lines, and their relative intensities match well. For lines A1, A5, and A6, in CRM simulation A1 is the strongest while in experiment A5 is. The possible reasons could be the uncertainties of the calculated transition rates and the neglected dynamical processes in the CRM simulation such as charge exchange. In addition, the efficiency correction may cause the discrepancy, since the efficiency curves given by the factory may also have some uncertainties.

Considering the limited number of configurations in our small-scale CRM simulation, we replace $\lambda_{\text{FAC}}^{\text{a}}$ by those obtained with our large-scale FAC calculations $\lambda_{\text{FAC}}^{\text{b}}$ and GRASP calculations $\lambda_{\text{GRA}}$ (Table I). The better agreement between our experimental and theoretical results is found for the wavelengths. Fig. 4 shows our merged theoretical simulations (wavelengths from our large-scale FAC and GRASP calculations, with intensities from our CRM simulation with small-scale FAC calculation) with a synthetic spectrum obtained from our experimental results. As can be seen, the average deviations of six lines from experiments with our large-scale FAC calculations $D_{\text{FAC}}^{\text{b}}$ and GRASP calculations $D_{\text{GRA}}$ are 0.88% and 0.83% respectively. The improved difference compared with our small-scale calculation indicate the significance of considering more electron correlations.

TABLE III. Identification for the experimentally observed lines based on our CRM simulation. Level numbers in the second column are in accordance with Table I. $\lambda_{\text{FAC}}^{\text{a}}$ (in nm) represents the wavelength from our small-scale FAC calculation in our CRM simulation. $\lambda_{\text{FAC}}^{\text{b}}$ and $\lambda_{\text{GRA}}$ represent wavelengths (in nm) of these six lines from our large-scale FAC ($\lambda_{\text{FAC}}^{\text{b}}$) and GRASP ($\lambda_{\text{GRA}}$) calculations deduced from Table I. $D_{\text{FAC}}^{\text{a}}$, $D_{\text{FAC}}^{\text{b}}$ and $D_{\text{GRA}}$ are the corresponding difference between our calculations and our experiments. $I_{\text{CRM}}$ is the intensity given by CRM, which is normalized to line A1. Transition rates (in s$^{-1}$) by our FAC ($A_{\text{FAC}}$) and GRASP ($A_{\text{GRA}}$) calculations are also presented. The numbers in square brackets stand for the power of 10.

| Key | Transition | $\lambda_{\text{FAC}}^{\text{a}}$ | $\lambda_{\text{FAC}}^{\text{b}}$ | $\lambda_{\text{GRA}}$ | $D_{\text{FAC}}^{\text{a}}$ | $D_{\text{FAC}}^{\text{b}}$ | $D_{\text{GRA}}$ | $I_{\text{CRM}}$ | $A_{\text{FAC}}$ | $A_{\text{GRA}}$ |
|---|---|---|---|---|---|---|---|---|---|---|
| A1 | 9→4 | 425.15 | 429.40 | 430.56 | -1.55 | -0.57 | -0.30 | 1.00 | 1.78[2] | 1.74[2] |
| A2 | 19→13 | 440.43 | 445.98 | 439.64 | -1.52 | -0.28 | 1.70 | 0.23 | 1.60[2] | 1.63[2] |
| A3 | 8→3 | 468.02 | 471.49 | 479.35 | -1.96 | -1.23 | 0.41 | 0.35 | 9.99[1] | 9.09[1] |

| A4 | 17→12 | 558.61 | 559.62 | 572.84 | -2.12 | -1.94 | 0.37 | 0.16 | 8.61[1] | 8.53[1] |
| A5 | 13→11 | 616.79 | 614.04 | 615.15 | 0.90 | 0.45 | 0.63 | 0.52 | 1.12[2] | 1.12[2] |
| A6 | 6→2 | 644.28 | 650.94 | 656.13 | -0.21 | 0.82 | 1.62 | 0.39 | 7.05[1] | 6.60[1] |

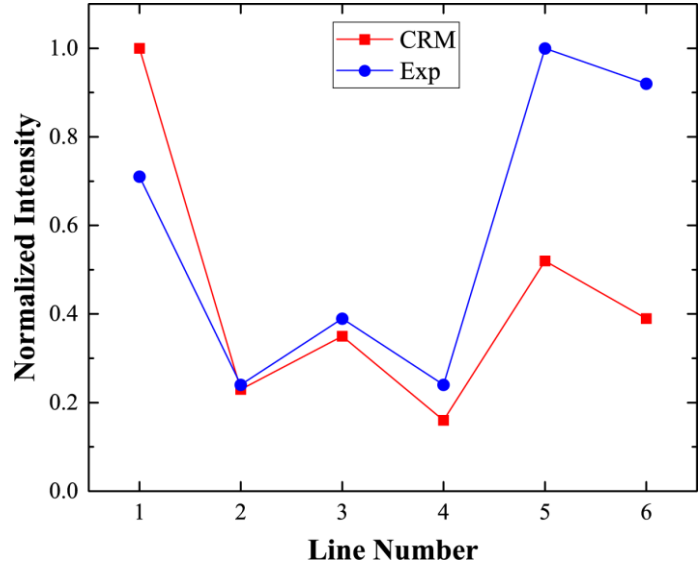

FIG. 3. Comparison between the intensities of six lines A1–A6 from experiments and our CRM simulation. Note, the experimental intensities are normalized to A5, while the CRM intensities are normalized to A1.

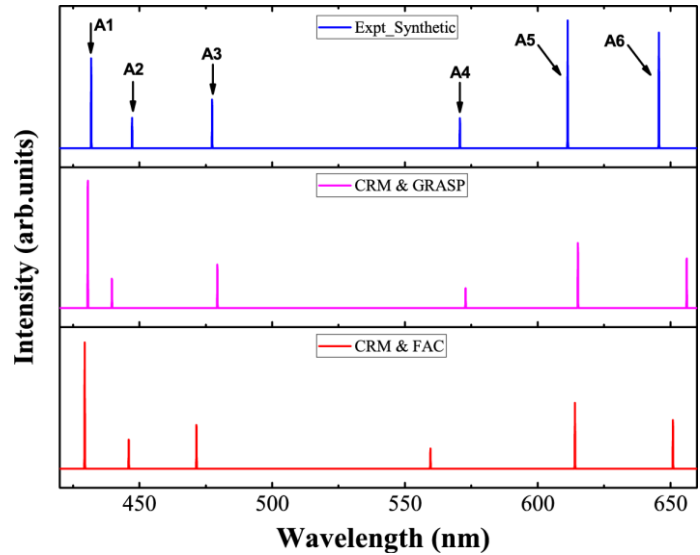

FIG. 4. Comparison between the experimental and calculated spectra of the $W^{8+}$ ion in the range of 420–660 nm. The top panel is the synthetic spectrum based on the wavelength and intensity measured by SH-HtscEBIT at electron-beam energy 104.5 eV. The middle panel is the simulated spectrum with wavelength from our GRASP calculations and intensity from our CRM simulation (electron density $5 \times 10^{10}$ cm$^{-3}$ and electron-beam energy 104.5 eV). The bottom panel is the simulated spectrum with wavelength from our large-scale FAC calculations and intensity from our CRM simulation. Note, each line is assumed to have the Gaussian profile with the full width at half maximum (FWHM) 0.14 nm, which corresponds to our experimental resolution.

## V. CONCLUSIONS

This paper presents both experimental and theoretical studies on the complicated spectrum of $W^{8+}$. The spectra of $W^{8+}$ are measured in the visible range at the SH-HtscEBIT during low electron-beam-energy operations. Atomic data for the lowest 30 energy levels of $W^{8+}$ are obtained by large-scale FAC calculations. Sophisticated GRASP calculations are also conducted to verify our FAC calculations. It is found that the average deviation of the excitation energies of the lowest 30 energy levels of $W^{8+}$ given by our two theoretical calculations is 1.42%. A detailed collisional-radiative model is constructed to help identify the six observed lines. Based on the reasonable agreement between our experimental and theoretical results, we assign these lines to the appropriate atomic transitions.

While this paper was under review, similar experimental and theoretical results for $W^{8+}$ have been published by Priti *et al*. [65]. Our experimental values are in good agreement with theirs for the wavelengths, and the same assignments to these lines are obtained based on individual calculations by using different computational models. However, one more strong visible line A6 at 645.48 nm is reported in the present work. In addition, our large-scale FAC calculations fit better with measurements (~0.88% deviation on average) than those reported in Ref. [65]. The more accurate theoretical results in our work also demonstrate the effective capture of electron correlations in complex atomic systems like $W^{8+}$, especially for the important influence related to the 4*d* and 4*f* electron correlation.

## ACKNOWLEDGMENTS

This work was supported by the National Natural Science Foundation of China under Grants No. 11974080, 11874090, 11934014, 11704076, U1732140, and was partly supported by the Chinese National Fusion Project for ITER No. 2015GB117000. Q. Lu is grateful to Alexander Kramida from National Institute of Standards and Technology for detailed explanation of their theoretical calculation on $W^{8+}$.